\documentclass[11pt, twoside, a4paper]{article}

\usepackage[width=165mm, left=20mm, bottom=30mm, top=25mm, foot=15mm, includefoot]{geometry}
\usepackage{titlesec}
\usepackage[hang,flushmargin]{footmisc}
\usepackage{times}
\usepackage[font=small, labelfont=it, textfont=it, labelsep=period]{caption}
\usepackage{indentfirst}
\usepackage[longnamesfirst,authoryear]{natbib}
\usepackage{amsmath, amsfonts, bbm}
\usepackage{graphicx}
\usepackage{color, bm}
\usepackage[english]{babel}
\usepackage[hidelinks]{hyperref}
\usepackage{booktabs}
\usepackage{threeparttable}

\usepackage{comment}
\usepackage{appendix}
\usepackage{amsthm}

\newcommand\blfootnote[1]{
  \begingroup
  \renewcommand\thefootnote{}\footnote{#1}
  \addtocounter{footnote}{-1}
  \endgroup
}

\titleformat{\section}
       {\normalfont\fontsize{11}{11}\bfseries}{\thesection.}{0.25em}{}
\titleformat{\subsection}
       {\normalfont\fontsize{11}{11}\itshape}{\thesubsection}{0.25em}{}
\titleformat{\subsubsection}
       {\normalfont\fontsize{11}{11}}{\thesubsubsection}{0.25em}{}
\titlespacing\section{0pt}{22pt plus 4pt minus 2pt}{11pt plus 4pt minus 2pt}
\titlespacing\subsection{0pt}{22pt plus 4pt minus 2pt}{11pt plus 4pt minus 2pt}
\titlespacing\subsubsection{0pt}{22pt plus 4pt minus 2pt}{11pt plus 4pt minus 2pt}

\newtheorem{Proposition}{Proposition}[subsection]
\newtheorem*{Proposition321}{Proposition 3.2.1}
\newtheorem*{Proposition322}{Proposition 3.2.2}

\renewcommand{\d}{\mathrm{d}}

\begin{document}
\thispagestyle{empty}

\begin{flushright}
\large{\textbf{Paolo Onorati$^\star$}\blfootnote{$^\star$ MEMOTEF, Sapienza University of Rome}},
\large{\textbf{Brunero Liseo$^\dagger$}\blfootnote{$^\dagger$ MEMOTEF, Sapienza University of Rome}}
\end{flushright}
\smallskip
\begin{flushleft}
\huge{\textbf{A Random Number Generator for the Kolmogorov Distribution}}\\
\end{flushleft}
\vspace{44pt}

\noindent\footnotesize{\emph{Abstract.} We discuss an acceptance-rejection algorithm for the random number generation from the Kolmogorov distribution. Since the cumulative distribution function (CDF) is expressed as a series, in order to obtain the density function we need to prove that the series of the derivatives converges uniformly. We also provide a similar proof in order to show that the ratio between the target Kolmogorov density and the auxiliary density implemented is bounded. Finally we discuss a way of truncating the series expression of the density in an optimal way.}\\
\bigskip

\noindent\small{\emph{Keywords:} Acceptance-Rejection algorithm, Uniform Convergence, Monte Carlo methods, Logistic distribution}

\section{Introduction}\label{sec1}
The Kolmogorov distribution naturally arises in the so-called Kolmogorov-Smirnov test, \cite{kolmogorov1933}, \cite{smirnov1939}. Here we briefly describe the case of the so-called one sample test. Let

\begin{equation*}
	\begin{aligned}
			X_1,X_2,\dots,X_n &\overset{i.i.d.}{\sim} F(\cdot) \, , \\
			\hat{F}_n(x) &= \frac{1}{n} \sum_{i=1}^{n} \mathbb{I}_{\{X_i \leq x\}}(x) \, ,
	\end{aligned}
\end{equation*}
\noindent
where $\mathbb{I}_{\{A\}}(x)$ is the indicator function of the set $A$, so $\hat{F}_n(\cdot)$ is the empirical CDF associated to the observed sample with size $n$. Let $F_0(\cdot)$ be an absolutely continuous probability distribution and

\begin{equation*}
	D_n = \sqrt{n} \, \sup \vert \hat{F}_n(x)-F_0(x) \vert	\, .
\end{equation*}

\cite{kolmogorov1933} prove that under the null hypothesis $F(\cdot)=F_0(\cdot)$ the following result holds,

\begin{equation*}
	\Lambda(x) = \lim\limits_{n\rightarrow+\infty} \mathrm{P} \left( D_n \leq x \right) = 
	\sum_{k=-\infty}^{+\infty} (-1)^k \exp(-2k^2x^2) \, , \quad x > 0  .
\end{equation*}
\noindent
Therefore the asymptotic distribution of $D_n$ is called Kolmogorov distribution and if the null hypothesis is true, it does not depend on $F_0(\cdot)$ as long as $F_0(\cdot)$ is absolutely continuous. \cite{smirnov1939} provided $2$ alternative and equivalent representations of the CDF of Kolmogorov distribution i.e.

\begin{equation*}
	\begin{aligned}
		\Lambda_1(x) &= 1-2\sum_{k=1}^{+\infty} (-1)^{k-1} \exp(-2k^2x^2) \, , \quad x > 0  , \\
		\Lambda_2(x) &= \frac{\sqrt{2\pi}}{x}\sum_{k=1}^{+\infty} \exp \left( -\frac{(2k-1)^2\pi^2}{8x^2} \right) \, , \quad x > 0  ,
	\end{aligned}
\end{equation*}
\noindent
where $\Lambda(x)=\Lambda_1(x)$ is easy to prove by simply algebraic manipulation and $\Lambda_1(x)=\Lambda_2(x)$ follows from the transformation formula for Theta functions, \cite{feller1948}, \cite{smirnov1939}.

The Kolmogorov distribution also arises in an other completely different framework i.e. in a representation of logistic distribution as a scale mixtures of Gaussian random variables. \cite{andrews1974} and \cite{stefanski1991} proved the following result; if

\begin{equation*}
	\begin{aligned}
		Y \vert W \sim& N(0, 4W^2) \, , \\
		W \sim& \Lambda(\cdot) \, ,
	\end{aligned}
\end{equation*}
then
\begin{equation*}
	Y \sim Logis(0,1) \, ,
\end{equation*}
\noindent
where $N(\mu, \sigma^2)$ denotes the normal distribution with mean $\mu$ and variance $\sigma^2$ and
$Logis(a,b)$ denotes the logistic distribution with density

\begin{equation*}
	\frac{ \exp( -\frac{x-a}{b} ) }{ b( 1+\exp(-\frac{x-a}{b}) )^2 } \, .
\end{equation*}

\subsection{Simulation from Kolmogorov distribution}\label{subsec1.1}
There is a stochastic representation of the Kolmogorov distribution: it is know that if $B(t)$ is a Brownian bridge and $X = \sup \vert B(t) \vert$, then $X \sim \Lambda(\cdot)$, \cite{perman2014}. This representation can be useful for the random number generation. However to generate a Brownian bridge and to get the supremum absolute value is computationally expensive, especially when one needs to generate several values from the Kolmogorov distribution.

Therefore in this paper we propose to use an acceptance-rejection algorithm. Let $f(x)$ be a target density with support $S$ and $g(x)$ an auxiliary density such that there exists a constant $M$ which verifies
\begin{equation*}
	\frac{f(x)}{Mg(x)} \le 1 \, , \quad x \in S \, , \quad 0 < M < +\infty \, .
\end{equation*}
Hence one can use the following procedure
\begin{enumerate}
	\item sample $x$ from $g(x)$,
	\item compute 
	\[
	p = \frac{f(x)}{Mg(x)} \, ,
	\]
	\item sample $u$ from an uniform distribution in $(0,1)$,
	\item if $u \le p$ get $x$ as a sample from $f(x)$.
\end{enumerate}
 Obviously in our case the target density is the Kolmogorov distribution. We use as auxiliary densities the Gamma and the inverse Gamma distributions. As we will explain, we obtain acceptance rates equals to $89.04\%$ and $95.23\%$ respectively.

\section{The density function}\label{sec2}
In the acceptance-rejection method we need to compute the density of Kolmogorov distribution. Since both representations of the CDF are in terms of series we need to prove that the series of derivatives converges uniformly in order to differentiate the series term by term. Notice that we say $\sum\limits_{k=1}^{+\infty} h_k(x)$ converges uniformly in $\mathrm{A} \subseteq \mathbb{R}$ if
\begin{equation*}
	\lim\limits_{n\rightarrow+\infty} \sup\limits_{x \in \mathrm{A}} \bigg\vert \sum\limits_{k=1}^{+\infty} h_k(x) - \sum\limits_{k=1}^{n} h_k(x) \bigg\vert = 0 \, ,
\end{equation*}
furthermore a sufficient condition for the uniform convergence is provided by Weierstrass criterion i.e. if
\begin{align*}
	\vert h_k(x) \vert \le& M_k \, , \quad x \in \mathrm{A} \, , \, k = 1,2,\dots \, \, \mathrm{and} \\
	\sum\limits_{k=1}^{+\infty}M_k <& +\infty \, ,
\end{align*}
then $\sum\limits_{k=1}^{+\infty} h_k(x)$ converges uniformly in $\mathrm{A}$.

First of all we note that
\begin{align}
	\frac{d}{dx}\left( (-1)^{k-1} \exp(-2k^2x^2) \right) &= (-1)^k 4k^2 x \exp(-2k^2 x^2) \, , \label{eq1}\\
	\frac{d}{dx} \exp \left( -\frac{(2k-1)^2\pi^2}{8x^2} \right) &= \frac{(2k-1)^2 \pi^2}{4x^3} \exp\left( -\frac{(2k-1)^2\pi^2}{8x^2} \right) \, , \label{eq2}
\end{align}
so from (\ref{eq1}) we require to prove the following proposition.
\begin{Proposition}\label{prop1}
	Let $ \mathrm{A} = \{ x \in \mathbb{R}: x \ge x_0 \}$ for some $x_0 > 0$ and
	\begin{equation*}
		h_k(x) = (-1)^k 4k^2 x \exp(-2k^2 x^2) \, ,
	\end{equation*}
	then $\sum\limits_{k=1}^{+\infty} h_k(x)$
	converges uniformly in $\mathrm{A} \, $.
\end{Proposition}
{\bf Proof:}
Let
\begin{equation*}
	a_k(x) =  4k^2 x \exp(-2k^2 x^2) \, ,
\end{equation*}
hence $h_k(x) = (-1)^k a_k(x)$. We fix $x$, with $x \ge x_0 > 0$, so $\{h_k(x)\}_{k=1}^{+\infty}$ is an alternating sequence with $a_k(x) > 0$; it is easy to show that
\begin{equation*}
	a_k(x) < a_{k+1}(x) \quad \mathrm{if} \quad k>1/(x\sqrt{2})
\end{equation*}
and
\begin{equation*}
	\lim\limits_{k\rightarrow+\infty}a_k(x)=0 \, .
\end{equation*}
Therefore the series $\sum\limits_{k=1}^{+\infty} h_k(x)$ converges point-wise by Leibniz criterion and we know
\begin{equation*}
	\bigg\vert \sum\limits_{k=1}^{+\infty} h_k(x) - \sum\limits_{k=1}^{n} h_k(x) \bigg\vert \le a_{n+1}(x) \, ,
\end{equation*} 
which implies
\begin{equation} \label{eq3}
	\sup\limits_{x \ge x_0} \bigg\vert \sum\limits_{k=1}^{+\infty} h_k(x) - \sum\limits_{k=1}^{n} h_k(x) \bigg\vert \le \sup\limits_{x \ge x_0} a_{n+1}(x) \, . 
\end{equation} 
Hence it is easy to compute
\begin{equation*}
	a'_{n+1}(x) = 4(n+1)^2 \exp\left( -2(n+1)^2x^2 \right) \left( 1-4x^2(n+1)^2\right) \, ,
\end{equation*}
so a global maximum exists for $x = \big(2(n+1)\big)^{-1}$, but we have restricted the support to $x \ge x_0$. Thus
\begin{equation*}
	\arg \, \max\limits_{x \ge x_0} \, a_{n+1}(x) = \left\{
	\begin{alignedat}{2}
		&\big(2(n+1)\big)^{-1} \quad &\mathrm{if} \quad \big(2(n+1)\big)^{-1} > x_0 \\
		& \ x_0 \quad &\mathrm{if} \quad \big(2(n+1)\big)^{-1} \le x_0
	\end{alignedat}
    \right. \ ,
\end{equation*}
so we have 
\begin{equation*}
	\max\limits_{x \ge x_0} \, a_{n+1}(x) = \left\{
	\begin{alignedat}{2}
		&2(n+1) \exp \left( -\frac{1}{2} \right) \quad &\mathrm{if} \quad \big(2(n+1)\big)^{-1} > x_0 \\
		&4(n+1)^2 x_0 \exp \left( -2(n+1)^2 x_0^2 \right) \quad &\mathrm{if} \quad \big(2(n+1)\big)^{-1} \le x_0
	\end{alignedat}
	\right. \ .
\end{equation*}
It is straightforward to see that
\begin{equation*}
	\lim\limits_{n\rightarrow+\infty} \arg \, \max\limits_{x \ge x_0} \, a_{n+1}(x) = x_0 \, ,
\end{equation*}
so, by taking the limit of (\ref{eq3}) we obtain
\begin{equation*}
	\lim\limits_{n\rightarrow+\infty} \sup\limits_{x \ge x_0} \bigg\vert \sum\limits_{k=1}^{+\infty} h_k(x) - \sum\limits_{k=1}^{n} h_k(x) \bigg\vert \le \lim\limits_{n\rightarrow+\infty} 4(n+1)^2 x_0 \exp \left( -2(n+1)^2 x_0^2 \right) = 0 \, .
\end{equation*}
Notice that if we extend the domain set $\mathrm{A}$ of \textbf{Proposition \ref{prop1}} from $x \ge x_0$ to $x \ge 0$, then we have
\begin{align*}
	\arg \, \max\limits_{x\ge0} \, a_{n+1}(x) &= \frac{1}{2(n+1)} \, , \\
	\max\limits_{x\ge0} \, a_{n+1}(x) &= 2(n+1) \exp \left( -\frac{1}{2} \right) \, .
\end{align*}
However in this case the limit for $n\rightarrow+\infty$ is not $0$. Therefore if we set the lower bound of $\mathrm{A}$ equal to $0$, the sufficient condition provided by Leibniz criterion fails.

Now we need to prove the following proposition.
\begin{Proposition}\label{prop2}
	Let $ \mathrm{A} = \{ x \in \mathbb{R}: 0 < x \le x_0 \}$ for some $x_0 > 0$ and
	\begin{equation*}
		h_k(x) = \frac{(2k-1)^2 \pi^2}{4x^3} \exp\left( -\frac{(2k-1)^2\pi^2}{8x^2} \right) \, ,
	\end{equation*}
	then $\sum\limits_{k=1}^{+\infty} h_k(x)$
	converges uniformly in $\mathrm{A} \, $.
\end{Proposition}
{\bf Proof:}
We use the sufficient condition provided by Weierstrass criterion, so we set
\begin{equation*}
	M_k = \max\limits_{0 < x \le x_0} \vert h_k(x) \vert = \max\limits_{0 < x \le x_0} h_k(x) \, ,
\end{equation*}
since $h_k(x)$ is always positive for $x>0$. It is easy to show that
\begin{equation*}
	h'_k(x) = \frac{(2k-1)^2\pi^2}{4x^4} \left( \frac{(2k-1)^2\pi^2}{4x^2} - 3 \right) \exp \left( - \frac{(2k-1)^2 \pi^2}{8x^2} \right) \, ,
\end{equation*}
hence a global maximum exists for $x = (2k-1)\pi / \sqrt{12}$, but we have restricted the space to $0 < x \le x_0$ thus
\begin{equation*}
	\arg \, \max\limits_{0 < x \le x_0} \, h_{k}(x) =
	\left\{
	\begin{alignedat}{2}
		&(2k-1)\pi / \sqrt{12} \quad &\mathrm{if} \quad (2k-1)\pi / \sqrt{12} < x_0 \\
		& \ x_0 \quad &\mathrm{if} \quad (2k-1)\pi / \sqrt{12} \ge x_0
	\end{alignedat}
	\right. \ ,
\end{equation*}
so we have 
\begin{equation*}
	\max\limits_{0 < x \le x_0} \, h_{k}(x) = M_k = 
	\left\{
	\begin{alignedat}{2}
		&\frac{2\sqrt{27}}{(2k-1)\pi} \exp \left( -\frac{3}{2} \right) \quad &\mathrm{if} \quad (2k-1)\pi / \sqrt{12} < x_0 \\
		& \frac{(2k-1)^2 \pi^2}{4x^3} \exp \left( -\frac{(2k-1)^2 \pi^2}{8x^2} \right) \quad &\mathrm{if} \quad (2k-1)\pi / \sqrt{12} \ge x_0
	\end{alignedat}
	\right. \ ,
\end{equation*}
and it is easy to prove that $\sum\limits_{k=1}^{+\infty} M_k$ is finite.

Notice that if we extend the domain set $\mathrm{A}$ of \textbf{Proposition \ref{prop2}} from $0 < x \le x_0$ to $\, 0 < x < +\infty$ then we have
\begin{align*}
	\arg \, \max\limits_{0 < x < +\infty} \, h_{k}(x) &= \frac{(2k-1)\pi}{\sqrt{12}} \, , \\
		\max\limits_{0 < x < +\infty} \, h_{k}(x) = M_k &= \frac{2\sqrt{27}}{(2k-1)\pi} \exp \left( -\frac{3}{2} \right) \, .
	\end{align*}
However in this case the series $\sum\limits_{k=1}^{+\infty} M_k$ is divergent. Therefore, if the definition set is unbounded above, the sufficient condition provided by Weierstrass criterion fails.

Finally, we obtain the density by deriving the series term by term. Let $\Lambda_1'(x) = \lambda_1(x)$ and $\Lambda_2'(x) = \lambda_2(x)$; we get for some $x_0 > 0$,
\begin{align*}
	\lambda_1(x) &= 8x \sum\limits_{k=1}^{+\infty} (-1)^{k-1} k^2 \exp \left(-2k^2 x^2 \right) \, , \, &x \ge x_0 \, , \\
	\lambda_2(x) &= \frac{\sqrt{2\pi}}{x^2} \sum\limits_{k=1}^{+\infty} \left( \frac{(2k-1)^2 \pi^2}{4x^2} -1 \right) \exp \left( -\frac{(2k-1)^2 \pi^2}{8x^2} \right) \, , \, &0 < x \le x_0 \, .
\end{align*}

\section{The proposal density}\label{sec3}
As the proposal density we explore the use of the Gamma and the inverse Gamma distributions, so we need to verify if the ratios between Kolmogorov density and the proposals are bounded. Notice that the target density is always finite for $0 < x < +\infty$; hence we only need to verify that the limit ratios are bounded as $x \rightarrow 0^+$ and $x \rightarrow +\infty$ that is
\begin{equation}\label{eq4}
	\lim\limits_{x \rightarrow +\infty} \frac{\lambda_1(x)}{g(x)} < +\infty \, , \, \lim\limits_{x \rightarrow 0^+} \frac{\lambda_2(x)}{g(x)} < +\infty \, ,
\end{equation}
where $g(x)$ is the proposal density. Notice that we cannot use
\begin{equation*}
	\lim\limits_{x \rightarrow +\infty} \frac{\lambda_2(x)}{g(x)} < +\infty \ \mathrm{or} \, \lim\limits_{x \rightarrow 0^+} \frac{\lambda_1(x)}{g(x)} < +\infty
\end{equation*}
because in section \ref*{sec2} we fail to prove that $\lambda_1(x)$ and $\lambda_2(x)$ are valid representations of Kolmogorov density as $x \rightarrow 0^+$ and $x \rightarrow +\infty$ respectively.

\subsection{Inverse Gamma proposal}\label{subsec3.1}
Let 
\begin{equation*}
	g(x) = \frac{\beta^\alpha}{\Gamma(\alpha)} x^{-\alpha-1} \exp \left( -\frac{\beta}{x} \right) \, , \quad x > 0 \, ,
\end{equation*}
hence by (\ref{eq4}) we must prove that
\begin{align}
	\label{eq6}
	&\lim\limits_{x \rightarrow +\infty} \frac{8\Gamma(\alpha)}{\beta^\alpha} \sum\limits_{k=1}^{+\infty} (-1)^{k-1} k^2 x^{\alpha+2} \exp \left(-2k^2 x^2 + \frac{\beta}{x}\right) < +\infty \, , \\
	\label{eq7}
	&\lim\limits_{x \rightarrow 0^+} \frac{\sqrt{2\pi}\Gamma(\alpha)}{\beta^\alpha} \sum\limits_{k=1}^{+\infty} \left( \frac{(2k-1)^2 \pi^2}{4x^2} -1 \right) x^{\alpha-1} \exp \left( -\frac{(2k-1)^2 \pi^2}{8x^2} + \frac{\beta}{x}\right) < +\infty \, .
\end{align}
In order to change the order between the limits and sums we need to prove that the series converges uniformly.
\begin{Proposition}\label{prop3}
		Let $ \mathrm{A} = \{ x \in \mathbb{R}: x \ge x_0 \}$ for some $x_0 > 0$ and
	\begin{equation*}
		h_k(x) = (-1)^{k-1} k^2 x^{\alpha+2} \exp \left(-2k^2 x^2 + \frac{\beta}{x}\right) \, ,
	\end{equation*}
	then $\sum\limits_{k=1}^{+\infty} h_k(x)$
	converges uniformly in $\mathrm{A} \, $.
\end{Proposition}
{\bf Proof:}
Let
\begin{equation*}
	a_k(x) =  k^2 x^{\alpha+2} \exp \left( -2k^2 x^2 + \frac{\beta}{x} \right) \, ,
\end{equation*}
hence $h_k(x) = (-1)^{k-1} a_k(x)$. We fix $x$, with $x \ge x_0 > 0$, so $\{h_k(x)\}_{k=1}^{+\infty}$ is an alternating sequence with $a_k(x) > 0$; it is easy to see that
\begin{equation*}
	a_k(x) < a_{k+1}(x) \quad \mathrm{if} \quad k>1/(x\sqrt{2})
\end{equation*}
and
\begin{equation*}
	\lim\limits_{k\rightarrow+\infty}a_k(x)=0 \, .
\end{equation*}
Therefore the series $\sum\limits_{k=1}^{+\infty} h_k(x)$ converges point-wise by Leibniz criterion and we know
\begin{equation*}
	\bigg\vert \sum\limits_{k=1}^{+\infty} h_k(x) - \sum\limits_{k=1}^{n} h_k(x) \bigg\vert \le a_{n+1}(x) \, ,
\end{equation*} 
which implies
\begin{equation} \label{eq5}
	\sup\limits_{x \ge x_0} \bigg\vert \sum\limits_{k=1}^{+\infty} h_k(x) - \sum\limits_{k=1}^{n} h_k(x) \bigg\vert \le \sup\limits_{x \ge x_0} a_{n+1}(x) \, . 
\end{equation}
We consider
\begin{align*}
	a_{n+1}(x) &= (n+1)^2 x^{\alpha+2} \exp \left( -2(n+1)^2 x^2 + \frac{\beta}{x} \right) \, , \\
	b_{n+1}(x) &= (n+1)^2 x^{\alpha+2} \exp \left( -2(n+1)^2 x^2 + \frac{\beta}{x_0} \right) \, ,
\end{align*}
so $a_{n+1}(x) \le b_{n+1}(x)$ since $x \ge x_0$. 
Furthermore
\begin{equation*}
	\frac{\d \log b_{n+1}(x)}{ \d x} = \frac{\alpha+2}{x} - 4(n+1)^2 x \, 
\end{equation*}
and
\begin{equation*}
	\frac{\d \log b_{n+1}(x)}{ \d x} \ge 0
	\iff
	x^2 \le \frac{\alpha+2}{4(n+1)^2}
	\, ,
\end{equation*}
hence we obtain
\begin{equation*}
	\arg \, \max\limits_{x \ge x_0} b_{n+1}(x) =
	\left\{
	\begin{alignedat}{2}
		& \frac{ \sqrt{\alpha+2} }{2(n+1)} \quad &\mathrm{if} \quad x_0 < \frac{ \sqrt{\alpha+2} }{2(n+1)} \\
		& \ x_0 \quad &\mathrm{if} \quad x_0 \ge \frac{ \sqrt{\alpha+2} }{2(n+1)}
	\end{alignedat}
	\right. \ ,
\end{equation*}
It is straightforward 
\begin{align*}
	&\lim\limits_{n\rightarrow+\infty} \arg \, \max\limits_{x \ge x_0} \, b_{n+1}(x) = x_0 \, , \\
	&\lim\limits_{n\rightarrow+\infty} \max\limits_{x \ge x_0} \, b_{n+1}(x) = \lim\limits_{n \rightarrow +\infty} (n+1)^2 x_0^{\alpha+2} \exp \left( -2(n+1)^2 x_0^2 + \frac{\beta}{x_0} \right) = 0 \, ,
\end{align*}
so by taking the limit of (\ref{eq5}) we obtain
\begin{equation*}
	\lim\limits_{n\rightarrow+\infty} \sup\limits_{x \ge x_0} \bigg\vert \sum\limits_{k=1}^{+\infty} h_k(x) - \sum\limits_{k=1}^{n} h_k(x) \bigg\vert \le \lim\limits_{n\rightarrow+\infty} \sup\limits_{x \ge x_0} a_{n+1}(x) \le \lim\limits_{n\rightarrow+\infty} \sup\limits_{x \ge x_0} b_{n+1}(x) = 0 \, .
\end{equation*}
Notice that by \textbf{Proposition \ref{prop3}}, we can easily compute (\ref{eq6}) simply by switching the order between limit and summation:
\begin{equation*}
	\frac{8\Gamma(\alpha)}{\beta^\alpha} \sum\limits_{k=1}^{+\infty} \lim\limits_{x \rightarrow +\infty} (-1)^{k-1} k^2 x^{\alpha+2} \exp \left(-2k^2 x^2 + \frac{\beta}{x}\right) = 0 \, .
\end{equation*}

\begin{Proposition}\label{prop4}
	Let $ \mathrm{A} = \{ x \in \mathbb{R}: 0 < x \le x_0 \}$ for some $x_0 > 0$ and
	\begin{equation*}
		h_k(x) = \sum\limits_{k=1}^{+\infty} \left( \frac{(2k-1)^2 \pi^2}{4x^2} -1 \right) x^{\alpha-1} \exp \left( -\frac{(2k-1)^2 \pi^2}{8x^2} + \frac{\beta}{x}\right) \, ,
	\end{equation*}
	then $\sum\limits_{k=1}^{+\infty} h_k(x)$
	converges uniformly in $\mathrm{A} \, $.
\end{Proposition}
{\bf Proof:}
Let
\begin{equation*}
	a_k(x) = \frac{(2k-1)^2 \pi^2}{4}x^{\alpha-3} \exp \left( -\frac{(2k-1)^2 \pi^2}{8x^2} + \frac{\beta}{x}\right) \, ,
\end{equation*}
thus $h_k(x) \le a_k(x)$ since $0 < x \le x_0$, we consider the exponential term of $a_k(x)$, i.e.
\begin{equation*}
	\exp \left( -\frac{(2k-1)^2 \pi^2}{8x^2} + \frac{\beta}{x}\right) = \exp \left( \frac{1}{x} \left( \beta -\frac{(2k-1)^2 \pi^2}{8x} \right) \right) \, ,
\end{equation*}
then we have
\begin{equation*}
	\beta -\frac{(2k-1)^2 \pi^2}{8x} > 0  \iff x > \frac{(2k-1)^2 \pi^2}{8 \beta} \, .
\end{equation*}
We have $2$ different cases; if $x_0 > (2k-1)^2 \pi^2/(8 \beta)$ then
\begin{equation*}
	\exp \left( \frac{1}{x} \left( \beta -\frac{(2k-1)^2 \pi^2}{8x} \right) \right) \le \exp \left( \frac{8 \beta}{(2k-1)^2 \pi^2} \left( \beta -\frac{(2k-1)^2 \pi^2}{8x} \right) \right) \, ,
\end{equation*}
if $x_0 \le (2k-1)^2 \pi^2/(8 \beta)$ then
\begin{equation*}
	\exp \left( \frac{1}{x} \left( \beta -\frac{(2k-1)^2 \pi^2}{8x} \right) \right) \le \exp \left( \frac{1}{x_0} \left( \beta -\frac{(2k-1)^2 \pi^2}{8x} \right) \right) \, ,
\end{equation*}
therefore, let 
\begin{align*}
	%&\tilde{k} = \inf \left\{ k \in \mathbb{N}_0 : x_0 \le \frac{(2k-1)^2 \pi^2}{8 \beta} \right\} \, , \\
	&b_k(x) = \left\{ 
	\begin{alignedat}{2}
		&\frac{(2k-1)^2 \pi^2}{4} x^{\alpha-3} \exp \left( \frac{\beta}{x_0} -\frac{(2k-1)^2 \pi^2}{8x_0 x} \right) \quad &\mathrm{if} \quad x_0 \le \frac{(2k-1)^2 \pi^2} {8 \beta} \\
		&\frac{(2k-1)^2 \pi^2}{4} x^{\alpha-3} \exp \left( \frac{8 \beta^2}{(2k-1)^2 \pi^2} - \frac{\beta}{x} \right) \quad &\mathrm{if} \quad x_0 > \frac{(2k-1)^2 \pi^2} {8 \beta}
	\end{alignedat}
	\right. \, ,
\end{align*}
thus $h_k(x) \le a_k(x) \le b_k(x)$. We use the sufficient condition provided by Weierstrass criterion, so we set
\begin{equation*}
	M_k = \max\limits_{0 < x \le x_0} b_k(x) \le \max\limits_{0 < x \le x_0} h_k(x) \, ,
\end{equation*}
it is easy to obtain
\begin{equation*}
	\frac{\d \log b_{k}(x)}{\d x} = \left\{
	\begin{alignedat}{2}
		&\frac{\alpha-3}{x} + \frac{(2k-1)^2 \pi^2}{8x_0 x^2} \quad &\mathrm{if} \quad x_0 \le \frac{(2k-1)^2 \pi^2} {8 \beta} \\
		&\frac{\alpha-3}{x} + \frac{\beta}{x^2} \quad &\mathrm{if} \quad x_0 > \frac{(2k-1)^2 \pi^2} {8 \beta}
	\end{alignedat}
\right. \, ,
\end{equation*}
we have $3$ different cases; if $\alpha \ge 3$ then 
\begin{align*}
	\arg \, \max\limits_{0 < x \le x_0} b_k(x) &= x_0 \, , \ k = 1,2, \dots \, ,\\
	\max\limits_{0 < x \le x_0} b_k(x) &= \left\{ 
	\begin{alignedat}{2}
		&\frac{(2k-1)^2 \pi^2}{4} x_0^{\alpha-3} \exp \left( \frac{\beta}{x_0}  -\frac{(2k-1)^2 \pi^2}{8x_0^2} \right) \quad &\mathrm{if} \quad x_0 \le \frac{(2k-1)^2 \pi^2} {8 \beta} \\
		&\frac{(2k-1)^2 \pi^2}{4} x_0^{\alpha-3} \exp \left( \frac{8 \beta^2}{(2k-1)^2 \pi^2} -\frac{\beta}{x_0} \right) \quad &\mathrm{if} \quad x_0 > \frac{(2k-1)^2 \pi^2} {8 \beta}
	\end{alignedat}
	\right. \, ,
\end{align*}
if $\alpha < 3$ and $\beta \le (3-\alpha)x_0$ then
\begin{align*}
	\arg \, \max\limits_{0 < x \le x_0} b_k(x) &= \left\{ 
	\begin{alignedat}{3}
		& \ x_0 \quad &\mathrm{if} \quad x_0 \le \frac{(2k-1) \pi} {2\sqrt{2(3-\alpha)}}& \\
		&\frac{(2k-1)^2 \pi^2}{(3-\alpha) 8x_0} \quad &\mathrm{if} \quad \frac{(2k-1) \pi} {2\sqrt{2(3-\alpha)}} < x_0& \le \frac{(2k-1)^2 \pi^2} {8 \beta} \\
		& \frac{\beta}{3-\alpha} \quad &\mathrm{if} \quad x_0 > \frac{(2k-1)^2\pi^2}{8\beta}&
	\end{alignedat}
	\right. \, , \\
	\max\limits_{0 < x \le x_0} b_k(x) &= \left\{ 
	\begin{alignedat}{3}
		&\frac{(2k-1)^2 \pi^2}{4} x_0^{\alpha-3} \exp \left( \frac{\beta}{x_0}  -\frac{(2k-1)^2 \pi^2}{8x_0^2} \right) \quad &\mathrm{if} \quad x_0 \le \frac{(2k-1) \pi} {2\sqrt{2(3-\alpha)}}& \\  
		&\frac{ (2k-1)^{2\alpha-4} \, \pi^{2\alpha-4} }{(3-\alpha)^{\alpha-3} \, 2^{3\alpha-7} \, x_0^{\alpha-3}} \exp \left( \frac{\beta}{x_0} + \alpha - 3 \right) \quad &\mathrm{if} \quad \frac{(2k-1) \pi} {2\sqrt{2(3-\alpha)}} < x_0& \le \frac{(2k-1)^2 \pi^2} {8 \beta} \\
		&\frac{(2k-1)^2 \pi^2}{4} \left(\frac{\beta}{3-\alpha}\right)^{\alpha-3} \exp \left( \frac{8\beta^2}{(2k-1)^2 \pi^2} + \alpha -3 \right) \quad &\mathrm{if} \quad x_0 > \frac{(2k-1)^2 \pi^2} {8 \beta}&  
	\end{alignedat}
	\right. \, ,
\end{align*}
and if $\alpha < 3$ and $\beta > (3-\alpha)x_0$ then
\begin{align*}
	\arg \, \max\limits_{0 < x \le x_0} b_k(x) &= \left\{ 
	\begin{alignedat}{2}
		& \ x_0 \quad &\mathrm{if} \quad x_0 \le  \frac{(2k-1)^2 \pi^2} {8 \beta} \\
		& \frac{\beta}{3-\alpha} \quad &\mathrm{if} \quad x_0 > \frac{(2k-1)^2\pi^2}{8\beta}
	\end{alignedat}
	\right. \, , \\
	\max\limits_{0 < x \le x_0} b_k(x) &= \left\{ 
	\begin{alignedat}{2}
		&\frac{(2k-1)^2 \pi^2}{4} x_0^{\alpha-3} \exp \left( \frac{\beta}{x_0}  -\frac{(2k-1)^2 \pi^2}{8x_0^2} \right) \quad &\mathrm{if} \quad x_0 \le \frac{(2k-1)^2 \pi^2} {8 \beta}& \\
		&\frac{(2k-1)^2 \pi^2}{4} \left(\frac{\beta}{3-\alpha}\right)^{\alpha-3} \exp \left( \frac{8\beta^2}{(2k-1)^2 \pi^2} + \alpha -3 \right) \quad &\mathrm{if} \quad x_0 > \frac{(2k-1)^2 \pi^2} {8 \beta}&  
	\end{alignedat}
	\right. \, .
\end{align*}
It is straightforward to see that, for all $3$ cases, there exists $\tilde k$ such that for $k \ge \tilde k$ one has
\begin{align*}
	\arg \, &\max\limits_{0 < x \le x_0} b_k(x) = x_0 \, ,\\
	&\max\limits_{0 < x \le x_0} b_k(x) = \frac{(2k-1)^2 \pi^2}{4} x_0^{\alpha-3} \exp \left( \frac{\beta}{x_0}  -\frac{(2k-1)^2 \pi^2}{8x_0^2} \right) \, ,
\end{align*}
so it is easy to show
\begin{equation*}
	\sum\limits_{k=1}^{+\infty}M_k = \sum\limits_{k=1}^{\tilde k -1}M_k + \sum\limits_{k = \tilde k}^{+\infty}M_k < +\infty \, .
\end{equation*}
Notice that from \textbf{Proposition \ref{prop4}} we can easily compute (\ref{eq7}) simply by switching the order between limit and summation:
\begin{equation*}
	\frac{\sqrt{2\pi}\Gamma(\alpha)}{\beta^\alpha} \sum\limits_{k=1}^{+\infty} \lim\limits_{x \rightarrow 0^+} \left( \frac{(2k-1)^2 \pi^2}{4x^2} -1 \right) x^{\alpha-1} \exp \left( -\frac{(2k-1)^2 \pi^2}{8x^2} + \frac{\beta}{x}\right) = 0 \, .
\end{equation*}
Therefore the inverse Gamma distribution is an admissible proposal density for all values of $\alpha, \beta$ and $x_0$.

\subsection{Gamma proposal}\label{subsec3.2}
Let	
\begin{equation*}
	g(x) = \frac{\beta^\alpha}{\Gamma(\alpha)} x^{\alpha-1} \exp \left( -\beta x \right) \, , \quad x > 0 \, ,
\end{equation*}
hence by (\ref{eq4}) we must prove
\begin{align}
	\label{eq8}
	&\lim\limits_{x \rightarrow +\infty} \frac{8\Gamma(\alpha)}{\beta^\alpha} \sum\limits_{k=1}^{+\infty} (-1)^{k-1} k^2 x^{2-\alpha} \exp \left(-2k^2 x^2 + \beta x \right) < +\infty \, , \\
	\label{eq9}
	&\lim\limits_{x \rightarrow 0^+} \frac{\sqrt{2\pi}\Gamma(\alpha)}{\beta^\alpha} \sum\limits_{k=1}^{+\infty} \left( \frac{(2k-1)^2 \pi^2}{4x^2} -1 \right) x^{-\alpha-1} \exp \left( -\frac{(2k-1)^2 \pi^2}{8x^2} + \beta x \right) < +\infty \, .
\end{align}
As for the inverse Gamma, in order to change the order between the limits and sums we need to prove that the series converge uniformly.

\begin{Proposition}\label{prop5}
	Let $ \mathrm{A} = \{ x \in \mathbb{R}: x \ge x_0 \}$ for some $x_0 > 0$ and
	\begin{equation*}
		h_k(x) = (-1)^{k-1} k^2 x^{2-\alpha} \exp \left(-2k^2 x^2 + \beta x \right) \, ,
	\end{equation*}
	then $\sum\limits_{k=1}^{+\infty} h_k(x)$
	converges uniformly in $\mathrm{A} \, $.
\end{Proposition}
{\bf Proof:} See Appendix.

From \textbf{Proposition \ref{prop5}} we can easily compute (\ref{eq8}) simply by switching the order between limit and summation:
\begin{equation*}
	\frac{8\Gamma(\alpha)}{\beta^\alpha} \sum\limits_{k=1}^{+\infty} \lim\limits_{x \rightarrow +\infty} (-1)^{k-1} k^2 x^{2-\alpha} \exp \left(-2k^2 x^2 + \beta x \right) = 0 \, .
\end{equation*}

\begin{Proposition}\label{prop6}
	Let $ \mathrm{A} = \{ x \in \mathbb{R}: 0 < x \le x_0 \}$ for some $x_0 > 0$ and
	\begin{equation*}
		h_k(x) = \left( \frac{(2k-1)^2 \pi^2}{4x^2} -1 \right) x^{-\alpha-1} \exp \left( -\frac{(2k-1)^2 \pi^2}{8x^2} + \beta x \right) \, ,
	\end{equation*}
	then $\sum\limits_{k=1}^{+\infty} h_k(x)$
	converges uniformly in $\mathrm{A} \, $.
\end{Proposition}
{\bf Proof:} See Appendix.

Hence  by \textbf{Proposition \ref{prop6}} we can easily compute (\ref{eq9}) simply by switching the order between limit and summation:
\begin{equation*}
	\frac{\sqrt{2\pi}\Gamma(\alpha)}{\beta^\alpha} \sum\limits_{k=1}^{+\infty} \lim\limits_{x \rightarrow 0^+} \left( \frac{(2k-1)^2 \pi^2}{4x^2} -1 \right) x^{-\alpha-1} \exp \left( -\frac{(2k-1)^2 \pi^2}{8x^2} + \beta x \right) = 0 \, .
\end{equation*}
Therefore the Gamma distribution is an admissible proposal density for all values of $\alpha, \beta$ and $x_0$.

\section{Numerical Optimization}
Once obtained a representation for the Kolmogorov distribution density and admissible proposals for acceptance-rejection algorithm, we need to tune some parameters of the proposal densities for applications. In particular we need a value $k^*$ in order to truncate the series of $\lambda_1(\cdot)$ and $\lambda_2(\cdot)$, a value $x_0$ and values for the parameters $\alpha$ and $\beta$ of the proposals.

According to the accuracy of the computing machine, there exists a value for $\bar x$ such that $\exp(-x)$ is set equal to $0$ for all $x$ greater than $\bar x$. Therefore we define 
\begin{align*}
	\bar k_{1}(x) &= \sup \{k \in \mathbb{N}_0: 2k^2x^2 \le \bar x \} = \left\lfloor \frac{\sqrt{2 \bar{x} }}{2x} \, \right\rfloor \, , \\
	\bar k_{2}(x) &= \sup \{ k \in \mathbb{N}_0: (2k-1)^2 \pi^2 / (8x^2)  \le \bar x  \} = \left\lfloor \frac{x \sqrt{2 \bar{x}} }{\pi} + \frac{1}{2} \right\rfloor \, ,
\end{align*}
where $\lfloor \cdot \rfloor$ is the floor function. The values $\bar k_{1}(x)$ and $\bar k_{2}(x)$ depend on $x$; for computational reasons we avoid to calculate them for all values of $x$; since the far is decreasing in $x$ with $x \ge x_0$ and the latter is increasing in $x$ with $0 < x \le x_0$,  we use their maximum value by assuming $x = x_0$ i.e. we set
\begin{equation} \label{eq10}
	k^*(x_0) = \max \Big( \bar k_1(x_0) \ \bar k_2(x_0) \Big) \, .
\end{equation} 
Hence expression (\ref{eq10}) provides the value of $k^*$ as a function of $x_0$, so we choose
\begin{align*}
	x_0^* &= \inf \left\{ \arg \, \min\limits_{x_0 > 0} k^*(x_0)  \right\} \, , \\
	k^* &= k^*(x_0^*) \, .
\end{align*}
Finally we obtain a numerical approximation $f^*(\cdot)$ of the Kolmogorov density by
\begin{equation*}
	f^*(x) = \left\{ 
	\begin{alignedat}{2}
		&\frac{\sqrt{2\pi}}{x^2} \sum\limits_{k=1}^{k^*} \left( \frac{(2k-1)^2 \pi^2}{4x^2} -1 \right) \exp \left( -\frac{(2k-1)^2 \pi^2}{8x^2} \right) \quad &\mathrm{if} \quad 0 <& x < x_0^*	 \\
		&8x \sum\limits_{k=1}^{k^*} (-1)^{k-1} k^2 \exp \left(-2k^2 x^2 \right) \quad &\mathrm{if} \quad x \ge& x_0^*
	\end{alignedat}
	\right. \, .
\end{equation*}
In our machine the value of $\bar x$ is $745.13$ with an approximation of $2$ digits, so we obtain %$745.13321910194116$ with an approximation of $14$ digits
\begin{align*}
	x_0^* = 1.207 \, , \,  k^* = 15 \, .
\end{align*}

As for as the parameters of the proposal we proceed in the following way: let $g(\cdot; \alpha, \beta)$ be the proposal density depending, we set
\begin{equation*}
	M = h(\alpha, \beta) = \, \sup\limits_{x>0} \, \frac{f^*(x)}{g(x; \alpha, \beta)} \, ,
\end{equation*}
so we obtain
\begin{equation*}
	 (\alpha^*, \beta^*) = \arg \, \min\limits_{\alpha > 0, \beta > 0} h(\alpha, \beta) \, ,
\end{equation*}
where the minimization is computed via numerical approximation. For the inverse Gamma proposal we have $\alpha^* = 10.29$, $\beta^* = 8.33$ and $M = 1.05$; for the Gamma proposal we obtain $\alpha^* = 9.21$, $\beta^* = 10.96$ and $M = 1.123$. The values of $M$ lead to a theoretical acceptance rate $1/M$ equals to 
$95.23\%$ and $89.04\%$ for inverse Gamma and Gamma respectively.

Finally we perform a simulation study in order to evaluate the accuracy and efficiency of our procedure with the inverse Gamma proposal. We generate $100$ times $10^6$ draws; for each iteration the mean elapsed time was $1.3623 \, \mathrm{sec}$; furthermore we set the seed to $1$ and with a single iteration of $10^6$ draws we obtain an empirical mean and empirical variance equal to
\begin{equation*}
	\hat{ \mathrm{\mu} } = 8.687866 \times 10^{-1}  \, , \, \hat{ \mathrm{\sigma} }^2 = 6.76608 \times 10^{-2} \, ,
\end{equation*}
while, as reported in \cite{marsaglia2003}, the true values are:
\begin{equation*}
	\mu = \sqrt{\pi/2} \log 2 \approx 8.687311 \times 10^{-1} \, , \, \sigma^2 = \pi^2/12 - \mu^2 \approx 6.777320 \times 10^{-2} \, .
\end{equation*}

\section{Conclusion}\label{sec4}
In this paper we have derived the density function of the Kolmogorov distribution by deriving the series term by term; surprisingly the standard sufficient condition for the uniformly convergence fails to hold for both bounds of the domain set, so we have used $2$ different representations which uniformly converge in $1$ bound at time. Furthermore they are also useful to speed up the convergence of the series. We have provided an approximation of the obtained density function via truncation series based on the machine precision.

In the same way we have shown that Gamma and inverse Gamma distributions are admissible proposals for Kolmogorov distribution in acceptance-rejection algorithm and both of them provide very fine theoretical acceptance rate after numerical optimization of the parameters. A simulation study have shown the accuracy and efficiency of our method.

\begin{appendices}
	\section*{Appendix}
	Here we provide the proof of \textbf{ Proposition \ref{prop5} } and \textbf{ Proposition \ref{prop6} }.
	
	\begin{Proposition321}
		Let $ \mathrm{A} = \{ x \in \mathbb{R}: x \ge x_0 \}$ for some $x_0 > 0$ and
		\begin{equation*}
			h_k(x) = (-1)^{k-1} k^2 x^{2-\alpha} \exp \left(-2k^2 x^2 + \beta x \right) \, ,
		\end{equation*}
		then $\sum\limits_{k=1}^{+\infty} h_k(x)$
		converges uniformly in $\mathrm{A} \, $.
	\end{Proposition321}
	{\bf Proof:}
	Let
	\begin{equation*}
		a_k(x) =  k^2 x^{2-\alpha} \exp(-2k^2 x^2 + \beta x) \, ,
	\end{equation*}
	hence $h_k(x) = (-1)^{k-1} a_k(x)$. We fix $x$, with $x \ge x_0 > 0$, so $\{h_k(x)\}_{k=1}^{+\infty}$ is an alternating sequence with $a_k(x) > 0$, it is easy to show that
	\begin{equation*}
		a_k(x) < a_{k+1}(x) \quad \mathrm{if} \quad k>1/(x\sqrt{2})
	\end{equation*}
	and
	\begin{equation*}
		\lim\limits_{k\rightarrow+\infty}a_k(x)=0 \, .
	\end{equation*}
	Therefore the series $\sum\limits_{k=1}^{+\infty} h_k(x)$ converges point-wise by Leibniz criterion and we know
	\begin{equation*}
		\bigg\vert \sum\limits_{k=1}^{+\infty} h_k(x) - \sum\limits_{k=1}^{n} h_k(x) \bigg\vert \le a_{n+1}(x) \, ,
	\end{equation*} 
	which implies
	\begin{equation}\label{eq11}
		\sup\limits_{x \ge x_0} \bigg\vert \sum\limits_{k=1}^{+\infty} h_k(x) - \sum\limits_{k=1}^{n} h_k(x) \bigg\vert \le \sup\limits_{x \ge x_0} a_{n+1}(x) \, . 
	\end{equation}
	We define
	\begin{equation*}
		b_{n+1}(x) = \left\{
		\begin{alignedat}{2}
			&x_0^{2-\alpha} (n+1)^2 \exp \left( -2 (n+1)^2 x^2 + \beta x \right) \quad &\mathrm{if} \quad \alpha \ge 2 \\
			&a_{n+1}(x) \quad &\mathrm{if} \quad \alpha < 2
		\end{alignedat}
		\right. \, ,
	\end{equation*}
	so $a_{n+1}(x) \le b_{n+1}(x)$ since $x \ge x_0$, 
	furthermore
	\begin{equation*}
		\frac{\d \log b_{n+1}(x)}{ \d x} = \left\{
		\begin{alignedat}{2}
			&-4(n+1)^2x + \beta \quad &\mathrm{if} \quad \alpha \ge 2 \\
			&\frac{2-\alpha}{x} -4(n+1)^2x + \beta \quad &\mathrm{if} \quad \alpha < 2
		\end{alignedat}
		\right. \, .
	\end{equation*}
	Therefore, we have $2$ different cases, if $\alpha \ge 2$ then
	\begin{equation*}
		\arg \, \max\limits_{x \ge x_0} b_{n+1} = \left\{ 
		\begin{alignedat}{2}
			&\frac{\beta}{4(n+1)^2} \quad &\mathrm{if} \quad  \frac{\beta}{4(n+1)^2} \ge x_0 \\
			& \, x_0 \quad &\mathrm{if} \quad  \frac{\beta}{4(n+1)^2} < x_0
		\end{alignedat}
		\right. \, ,
	\end{equation*}
	if $\alpha < 2$ then 
	\begin{equation*}
		\arg \, \max\limits_{x \ge x_0} b_{n+1} = \left\{ 
		\begin{alignedat}{2}
			&\frac{ \beta + \sqrt{\beta^2 + 16(n+1)^2(2-\alpha)} }{8(n+1)^2} \quad &\mathrm{if} \quad  \frac{ \beta + \sqrt{\beta^2 + 16(n+1)^2(2-\alpha)} }{8(n+1)^2} \ge x_0 \\
			& \, x_0 \quad &\mathrm{if} \quad  \frac{ \beta + \sqrt{\beta^2 + 16(n+1)^2(2-\alpha)} }{8(n+1)^2} < x_0
		\end{alignedat}
		\right. \, ,
	\end{equation*}
	In both cases it is straightforward 
	\begin{align*}
		&\lim\limits_{n\rightarrow+\infty} \arg \, \max\limits_{x \ge x_0} \, b_{n+1}(x) = x_0 \, , \\
		&\lim\limits_{n\rightarrow+\infty} \max\limits_{x \ge x_0} \, b_{n+1}(x) = \lim\limits_{n \rightarrow +\infty} (n+1)^2 x_0^{2-\alpha} \exp \left( -2(n+1)^2 x_0^2 + \beta x_0 \right) = 0 \, ,
	\end{align*}
	so by taking the limit of (\ref{eq11}) we obtain
	\begin{equation*}
		\lim\limits_{n\rightarrow+\infty} \sup\limits_{x \ge x_0} \bigg\vert \sum\limits_{k=1}^{+\infty} h_k(x) - \sum\limits_{k=1}^{n} h_k(x) \bigg\vert \le \lim\limits_{n\rightarrow+\infty} \sup\limits_{x \ge x_0} a_{n+1}(x) \le \lim\limits_{n\rightarrow+\infty} \sup\limits_{x \ge x_0} b_{n+1}(x) = 0 \, .
	\end{equation*}
	
	\vspace{1.5pt}
	\begin{Proposition322}
		Let $ \mathrm{A} = \{ x \in \mathbb{R}: 0 < x \le x_0 \}$ for some $x_0 > 0$ and
		\begin{equation*}
			h_k(x) = \left( \frac{(2k-1)^2 \pi^2}{4x^2} -1 \right) x^{-\alpha-1} \exp \left( -\frac{(2k-1)^2 \pi^2}{8x^2} + \beta x \right) \, ,
		\end{equation*}
		then $\sum\limits_{k=1}^{+\infty} h_k(x)$
		converges uniformly in $\mathrm{A} \, $.
	\end{Proposition322}
    {\bf Proof:}
    Let
    \begin{equation*}
    	a_k(x) = \frac{(2k-1)^2 \pi^2}{4} x^{-\alpha-3} \exp \left( -\frac{(2k-1)^2 \pi^2}{8 x^2} + \beta x_0 \right) \, ,
    \end{equation*}
    thus
    \begin{equation*}
    	h_k(x) \le a_k(x) \, .
    \end{equation*}
     Hence 
     \begin{equation*}
     	\frac{ \d \log a_k(x)}{\d x} = -\frac{\alpha+3}{x} + \frac{(2k-1)^2 \pi^2}{4 x^3} \, , 
     \end{equation*}
     therefore
     \begin{align*}
     	\arg \; &\max_{0 < x \le x_0} a_k(x) = \left\{
     	\begin{alignedat}{2}
     		&\frac{(2k-1)\pi}{2 \sqrt{\alpha+3}} \quad &\mathrm{if} \quad \frac{(2k-1)\pi}{2 \sqrt{\alpha+3}} < x_0 \\
     		&x_0 \quad &\mathrm{if} \quad \frac{(2k-1)\pi}{2 \sqrt{\alpha+3}} \ge x_0
     	\end{alignedat}
     	\right. \, , \\
     	 &\max_{0 < x \le x_0} a_k(x) = \left\{
     	\begin{alignedat}{2}
     		&\frac{  (2k-1)^{-\alpha-1} \pi^{-\alpha-1}  }{  2^{-\alpha-1} (\alpha+3)^{ -\frac{\alpha+3}{2} }  } \exp \left( -\frac{\alpha+3}{2} + \beta x_0 \right) \quad &\mathrm{if} \quad \frac{(2k-1)\pi}{2 \sqrt{\alpha+3}} < x_0 \\
     		&\frac{(2k-1)^2 \pi^2}{4} x_0^{-\alpha-3} \exp \left( -\frac{(2k-1)^2 \pi^2}{8 x_0^2} + \beta x_0 \right) \quad &\mathrm{if} \quad \frac{(2k-1)\pi}{2 \sqrt{\alpha+3}} \ge x_0
     	\end{alignedat}
     \right. \, .
     \end{align*}
     So, let 
     \begin{align*}
     	\tilde k &= \inf \left\{ k \in \mathbb{N}_0: \frac{(2k-1)\pi}{2 \sqrt{\alpha+3}} \ge x_0 \right\} \, , \\
     	M_k &= \max_{0 < x \le x_0} a_k(x) \le a_k(x) \le h_k(x) \, ,
     \end{align*}
     we use the sufficient condition provided by Weierstrass criterion and it is easy to show
     \begin{equation*}
     	\sum\limits_{k=1}^{+\infty}M_k = \sum\limits_{k=1}^{\tilde k -1}M_k + \sum\limits_{k = \tilde k}^{+\infty}M_k < +\infty \, .
     \end{equation*}
    
\end{appendices}

\ifx\undefined\textsc
\newcommand{\textsc}[1]{{\sc #1}}
\newcommand{\emph}[1]{{\em #1\/}}
\let\tmpsmall\small
\renewcommand{\small}{\tmpsmall\sc}
\fi

\end{document}